\documentclass[11pt]{article}

\usepackage{graphicx}
\usepackage{slashed}
\begin{document}

\title{A Single Source for all Flavor Violation}
\author{{\bf S.M. Barr} and {\bf Heng-Yu Chen} \\
Department of Physics and Astronomy \\ 
and \\
Bartol Research Institute \\ University of Delaware
Newark, Delaware 19716} \maketitle

\begin{abstract}
In a model proposed in 2012, all flavor mixing has a single source
and is governed by a single ``master matrix." This model was shown to give 
several predictions for quark and lepton masses and mixing angles and
for mixing angles within $SU(5)$ multiplets that are observable
in proton decay. Here it is shown that the same master matrix controls 
the flavor-changing processes mediated by a singlet scalar that exists 
in the model, giving predictions for $\tau \rightarrow \mu \gamma$, 
$\tau \rightarrow e \gamma$, and 
$\mu \rightarrow e \gamma$.
 .  
\end{abstract}

\section{Introduction} 

In a 2012 paper \cite{BarrChen2012}, we proposed a model in which all flavor changing effects, including CKM mixing \cite{CKM} and MNS mixing \cite{MNS}, are controlled by a single ``master matrix."
In that paper, the model was shown to give several predictions for 
neutrino properties, including the Dirac neutrino CP phase, as well as 
post-dictions for quantities that are still not precisely known, such as the 
atmospheric and solar neutrino mixing angles, and $m_s/m_d$.  
In a subsequent paper \cite{BarrChen2013} it was pointed out that the same model
predicts all the mixing angles that come into gauge-boson-mediated
proton decay, thus giving further tests. 

In this paper, we show that the same model gives predictions for flavor-changing 
effects produced by the exchange of a Standard-Model-singlet scalar field that exists in
the model. In principle, therefore, certain parameters of the model could be measurable in three independent ways: by precise determination of neutrino and quark
properties, by proton decay branching ratios, and through flavor-changing decays such as $\tau \rightarrow \mu \gamma$, $\tau \rightarrow e \gamma$, and 
$\mu \rightarrow e \gamma$.

The model is based on two assumptions: (1) that $SU(5)$ symmetry relates quarks and leptons, and (2) that all flavor violation comes from mixing between three chiral fermion families that we shall denote  
${\bf 10}_i + \overline{{\bf 5}}_i$, $i=1,2,3$, and
$N$ vector-like fermion multiplets that we shall denote ${\bf 5}'_A +\overline{{\bf 5}}'_A$, 
$A=1,...,N$. ($N$ can be as small as 2.) In particular, it is assumed that the Yukawa terms that involve 
only the three chiral families are flavor-diagonal due to an abelian family symmetry group, which we shall call $G_F$. .
The vector-like fermion multiplets, on the other hand, do not transform under $K_F$, and as a consequence their mixing with the three chiral fermion families produces flavor violation among those families. More specifically, the mixing responsible for
flavor violation is between the $\overline{{\bf 5}}_i$ and the  
$\overline{{\bf 5}}'_A$ multiplets. This means that the model is of the ``lopsided" type
\cite{lopsided}.  Because the origin of flavor changing is in the $\overline{{\bf 5}}$
sector, it shows up more strongly in left-handed leptons (which are in
$\overline{{\bf 5}}$ multiplets) than in left-handed quarks (which are in
${\bf 10}$ multiplets). This gives a simple and elegant explanation of the fact that the MNS angles are much larger than the CKM angles, as is the basic idea of so-called ``lopsided models" \cite{lopsided}.

The effect of the mixing of $\overline{{\bf 5}}_i$ and the  
$\overline{{\bf 5}}'_A$ shows up in the effective low-energy theory of the known quarks and leptons as a $3 \times 3$ non-diagonal ``master matrix," which we call $A$, that appears in their mass matrices. By field and parameter re-definitions this master matrix can be brought to a simple triangular form (which we call $A_{\Delta}$), which contains only one complex and two real parameters, whose values can be completely determined from CKM mixing. This allows predictions for all other flavor-changing effects.

In section 2, we will review the model and show how it leads to predictions for flavor changing in the lepton sector and in proton decay. In section 3, we will analyze the flavor-changing effects that arise from the exchange of a Standard-Model-singlet scalar that exists in the model.

\section{Review of the model} 

We shall now review the model and its predictions for masses and mixing matrices. More details can be found at \cite{BarrChen2012,BarrChen2013}. The Yukawa terms of the model are 

\begin{equation}
\begin{array}{ccl}
{\cal L}_{Yuk} & = & Y_i ({\bf 10}_i {\bf 10}_i) \langle {\bf 5}_H
\rangle + y_i ({\bf 10}_i \overline{{\bf 5}}_i) \langle {\bf
5}^{\dag}_H \rangle
\\ & & \\
& + & \tilde{Y}_i ({\bf 10}_i {\bf 10}_i) \langle {\bf 45}_H \rangle
+ \tilde{y}_i ({\bf 10}_i \overline{{\bf 5}}_i) \langle {\bf
45}^{\dag}_H \rangle
\\ & & \\
& + & (\lambda_i/M_R)(\overline{{\bf 5}}_i \overline{{\bf
5}}_i) \langle {\bf 5}_H \rangle \langle {\bf 5}_H \rangle
\\ & & \\
& + &  Y'_{AB} ({\bf 5}_A \overline{{\bf 5}}_B) \langle {\bf 1}_H
\rangle + y'_{A i} ({\bf 5}_A \overline{{\bf 5}}_i) \langle {\bf
1}'_{Hi} \rangle,
\end{array}
\end{equation}

\noindent where the subscript $H$ denotes Higgs multiplets. Repeated indices ($i$, $A$, or $B$) are summed over. The first two lines contain typical Yukawa terms that give realistic quark and lepton masses. The third line is the effective Weinberg dim-5 operator that gives mass to the neutrinos in either the type-I or type-II see-saw mechanisms. All the terms in the first three lines involve only the multiplets ${\bf 10}_i + \overline{{\bf 5}}_i$ and are therefore flavor-diagonal. The fourth 
line of Eq. (1) contains the Yukawa terms that give mass to the vector-like fermions and mix 
$\overline{{\bf 5}}_i$ with $\overline{{\bf 5}}'_A$. These masses, coming from $SU(5)$-singlet Higgs fields, can be much larger than the weak scale, and indeed can 
even be of order the GUT scale. The only assumption that is required to fit the CKM and MNS mixing angles is that the masses generated by these two terms are of the {\it same} order, which we shall refer to as the ``heavy scale" $M_*$. This scale must be large enough to explain why these new fermions have not been observed. The fermions that do not get mass of order $M_*$, which consist of the ${\bf 10}_i$ and three linear combinations of  $\overline{{\bf 5}}_i$ and $\overline{{\bf 5}}'_A$, are the known quarks and leptons, which we will call the ``light fermions".  

There are many abelian family symmetries that could enforce the flavor-diagonal form of the terms in the first three lines of Eq. (1). A simple example (though not the simplest) is that given in \cite{BarrChen2012}, namely $G_F = K_1 \times K_2 \times K_3$, where (for a
given $i$ equal to 1, 2, or 3) $K_i$ is a $Z_2$ symmetry under which
${\bf 10}_i$, $\overline{{\bf 5}}_i$, and ${\bf 1}'_{Hi}$ are odd
and all other fields even. Note that the vacuum expectation values of ${\bf 1}'_{Hi}$ spontaneously break the abelian family symmetries; so that the last term in Eq. (1),
which mixes the $\overline{{\bf 5}}_i$ and $\overline{{\bf 5}}'_A$, does not respect 
the family symmetries and can give flavor violation.
It is important for the predictivity of the model that the
last two terms in Eq. (1) involve only $SU(5)$-singlet Higgs fields, as otherwise the ``master matrix" would be different for quarks and leptons. This can be ensured by another abelian symmetry that prevents the $SU(5)$ adjoint Higgs field from coupling in these terms \cite{BarrChen2012}.

The Yukawa terms in the first three lines of Eq. (1) give rise to the following diagonal mass matrices $m_u$, $m_d$, $m_{\ell}$, and $m_{\nu}$:

\begin{equation} 
u_i \; (m_u)_{ij} \; u^c_j,  \;\;\;\;\;
d_i \; (m_d)_{ij} \; d^c_j, \;\;\;\;\;
\ell_i \; (m_{\ell})_{ij} \; \ell^c_j, \;\;\;\;\;
\nu_i \; (m_{\nu})_j \; \nu_j
\end{equation}

\noindent These are not the mass matrices of the known fermions, because we have not yet taken into account the mixing of the $\overline{{\bf 5}}$ multiplets, of which $N$ linear combinations are ``heavy" and 3 are ``light". A block-diagonalization to separate the heavy and light $\overline{{\bf 5}}$ fermion states is needed in order to find the effective mass matrices for the light fermions.

Let us call $[Y'_{AB} \langle {\bf 1}_H \rangle] \equiv M_{AB}$ and $[y'_{Ai} \langle {\bf 1}'_{Hi} \rangle] \equiv \Delta_{Ai}$. Let us first examine the down-type quarks. 
These have a $(3+N) \times (3+N)$ mass matrix of the form

\begin{equation}
\left( d_{({\bf 10})}, D_{({\bf 5}')} \right)  
\left( \begin{array}{cc} m_d & 0 \\ & \\
\Delta & M \end{array} \right) \left( \begin{array}{c} d^c_{(\overline{{\bf 5}})} \\ \\
D^c_{(\overline{{\bf 5}}')} \end{array} \right),
\end{equation}

\noindent The block diagonalization is carried out by a bi-unitary transformation of the $(3 +N) \times (3 + N)$ mass matrix:

\begin{equation}
\left( \begin{array}{cc} m_d & 0 \\ & \\
\Delta & M \end{array} \right) \longrightarrow \underbrace{\left( \begin{array}{cc}
I & G^{\dag} \\ & \\ -G & I \end{array} \right)}_{\cong U_L^{\dag}} 
\left( \begin{array}{cc} m_d & 0 \\ & \\
\Delta & M \end{array} \right) \underbrace{\left( \begin{array}{cc} A & B \\ & \\
C & D \end{array} \right)}_{\cong U_R} = \left( \begin{array}{cc} M_d & 0 \\ & \\
0 & M'  \end{array} \right). 
\end{equation}

\noindent Here the elements of $G$ are small and $U_L$ is approximately diagonal, because the elements of $m_d$ are very small compared to those of $M$ and $\Delta$. One can give exact expressions for the matrices $A$, $B$, $C$, $D$, and $G$, which will be useful in section 3. Defining $T \equiv M^{-1} \Delta$, one can write

\begin{equation}
\begin{array}{ccl}
A & \equiv & [I + T^{\dag} T]^{-1/2}, \\ 
B & \equiv & [I + T^{\dag} T]^{-1/2} T^{\dag} = A T^{\dag} 
=  T^{\dag} [I +  T T^{\dag}]^{-1/2}
\equiv T^{\dag} D \\ 
C & \equiv & - T [I + T^{\dag} T]^{-1/2} = - T A =  -[I +  T T^{\dag}]^{-1/2} T \equiv - D T \\ 
D & \equiv & [I +  T T^{\dag}]^{-1/2} \\ 
G & \equiv & - M^{-1 \dag} D^{2 \dag} T m_d^{\dag}.
\end{array}
\end{equation}

\noindent  Since the elements of $\Delta$ and $M$ are of the same order, the elements of $T$ are of $O(1)$, and the matrices $A$, $B$, $C$, $D$ have off-diagonal elements of $O(1)$. By simply multiplying out Eq. (4) one sees that
the effective mass matrix of the three ``light" down-type quarks, namely $M_d$, is given by 

\begin{equation}
M_d = m_d A.
\end{equation}

\noindent The reason that $m_d$ gets multiplied on the right by the matrix $A$ is that the matrix $m_d$ originally appears in a term ${\bf 10}_i (m_d)_{ij} \overline{{\bf 5}}_j$, and the matrix $A$ represents the mixing of the $\overline{{\bf 5}}$ multiplets. The mass matrix $m_{\ell}$ of the charged leptons appears in a term $\overline{{\bf 5}}_i (m_{\ell})_{ij} {\bf 10}_j$, as can be seen from Eq. (2), and so gets multiplied on the left by $A^T$. Therefore, the effective mass matrix of the three light charged leptons is $M_{\ell} = A^T m_{\ell}$. The up quark mass matrix $m_u$ appears in a term ${\bf 10}_i (m_u)_{ij} {\bf 10}_j$, which involves no $\overline{{\bf 5}}$ multiplets, and so does not get multiplied by any factors of $A$. Therefore, $M_u = m_u$.  Finally, the mass matrix of the neutrinos, which comes from the dim-5 Weinberg effective 
operator, appears in a term $\overline{{\bf 5}}_i (m_{\nu})_{ij} \overline{{\bf 5}}_j$, and so gets multiplied on both the right by $A$ and the left by $A^T$. Hence, we have altogether

\begin{equation}
M_u = m_u, \;\;\;\;\;\; M_d = m_d A, \;\;\;\;\;\; 
M_{\ell} = A^T m_{\ell}, \;\;\;\;\;\;
M_{\nu} = A^T m_{\nu} A.
\end{equation}

\noindent We thus see that all flavor violation is controlled by $A$. Moreover, the matrix $A$ can be brought to a simple form in the following way. By multiplying $A$ on the right by a unitary matrix, the elements below the main diagonal of $A$ can be made zero. Then by rescaling the rows by multiplying from the left by a complex diagonal matrix, the diagonal elements of $A$ can be set to 1. That is, $A$ can be written 

\begin{equation}
A = {\cal D} \; A_{\Delta} \; {\cal U},
\end{equation}

\noindent where ${\cal D}$ is a complex diagonal matrix, ${\cal U}$ is a unitary matrix, and
$A_{\Delta}$ is a matrix of the form

\begin{equation}
A_{\Delta} = \left( \begin{array}{ccc} 1 & b & c e^{i \theta} \\ 0 & 1 & a \\ 0 & 0 & 1 \end{array}
\right),
\end{equation}

\noindent where $a$, $b$, and $c$ are real. It is easily seen that the matrix ${\cal U}$ can be absorbed into redefined right-handed down quarks and the left-handed lepton doublets. Similarly, the phases in ${\cal D}$ can be absorbed into redefined fields. The diagonal real matrix $|{\cal D}|$ can be absorbed into redefinitions of the original diagonal mass matrices as follows: $\overline{m}_d \equiv m_d |{\cal D}|$,
$\;\; \overline{m}_{\ell} \equiv m_{\ell} |{\cal D}|$, $\;\; \overline{m}_{\nu} 
\equiv m_{\nu} |{\cal D}|^2$, and $\overline{m}_u \equiv m_u$. Thus, after these redefinitions, the mass matrices of the three light families take the new form

\begin{equation}
\overline{M}_u = \overline{m}_u, \;\;\;\;\;\; \overline{M}_d = \overline{m}_d A_{\Delta}, \;\;\;\;\;\; 
\overline{M}_{\ell} = A_{\Delta}^T \overline{m}_{\ell}, \;\;\;\;\;\;
\overline{M}_{\nu} = A_{\Delta}^T \overline{m}_{\nu} A_{\Delta}.
\end{equation}

\noindent  It is easy to see that to a very good approximation the elements of the diagonal matrix $\overline{m}_d$ are just the eigenvalues of $\overline{M}_d$, {\it i.e.} the physical masses of the $d$, $s$, and $b$ quarks.  Therefore, in the basis of Eq. (10), the mass matrices of the up quarks and down quarks look as follows

\begin{equation}
\overline{M}_u = \left( \begin{array}{ccc} m_u & 0 & 0 \\ 0 & m_c & 0 \\ 0 & 0 & m_t \end{array}
\right); \;\;\;\;\;\;\;  \overline{M}_d \cong \left( \begin{array}{ccc} m_d & 0 & 0 \\ 0 & m_s & 0 \\ 0 & 0 & m_b \end{array} 
\right) \left( \begin{array}{ccc} 1 & b & c e^{i \theta} \\ 0 & 1 & a \\ 0 & 0 & 1 \end{array}
\right) = \left( \begin{array}{ccc} m_d & bm_d & c e^{i \theta} m_d \\ 0 & m_s & a m_s \\ 0 & 0 & m_b \end{array}
\right).
\end{equation}

\noindent One sees immediately that

\begin{equation}
\begin{array}{l}
|V_{cb}| \cong \frac{a m_s}{m_b}  \;\; \Longrightarrow \;\; a \cong \frac{m_b}{m_s} 
|V_{cb}| \sim 2, \\ \\
|V_{us}| \cong \frac{b m_d}{m_s} \;\; \Longrightarrow \;\;  b \cong \frac{m_s}{m_d} 
|V_{us}| \sim 4, \\ \\
V_{ub} \cong \frac{c e^{i \theta} m_d}{m_b} \;\; \Longrightarrow \;\; c \cong \frac{m_b}{m_d} |V_{ub}| \sim 4, \;\;\;\theta \cong \delta_{KM}.
\end{array}
\end{equation}

\noindent Similarly, the elements of the diagonal matrix $\overline{m}_{\ell}$ in Eq. (10) are to a very good approximation the masses of the $e$, $\mu$, and $\tau$. In the basis of Eq. (10), therefore, the charged lepton mass matrix $M_{\ell}$ has the form

\begin{equation}
\overline{M}_{\ell} \cong \left( \begin{array}{ccc} 1 & 0 & 0 \\
b & 1 & 0 \\ ce^{i \theta} & a & 1 \end{array} \right) \left( \begin{array}{ccc} m_e & 0 & 0 \\ 0 & m_{\mu} & 0 \\ 0 & 0 & m_{\tau} \end{array} 
\right) = \left( \begin{array}{ccc} m_e & 0 & 0 \\
b m_e & m_{\mu} & 0 \\ ce^{i \theta} m_e & a m_{\mu} & m_{\tau} \end{array}
\right).
\end{equation}

\noindent This is not diagonal, but the rotations required to diagonalize it are very small for {\it left-handed} charged leptons (namely, of $O(m_{\mu}^2/m_{\tau}^2)$, $O(m_{\mu} m_e/m_{\tau}^2)$, $O(m_e^2/m_{\tau}^2)$). Thus, to a very good approximation, in the basis where the charged lepton mass matrix is diagonal, the effective neutrino mass matrix $\overline{M}_{\nu}$ has the form (from Eqs. (9), (10) and (12))
 
\begin{equation}
\overline{M}_{\nu} \cong
\left[ \begin{array}{ccc} 1 & 0 & 0 \\
\frac{m_s}{m_d} | \overline{V}_{us}| & 1 & 0 \\
\frac{m_b}{m_d} |V_{ub}| e^{i \delta} & \frac{m_b}{m_s} |V_{cb}| & 1
\end{array} \right]
\left[ \begin{array}{ccc} q e^{i \beta} & 0 & 0
\\ 0 & p e^{i \alpha} & 0 \\
0 & 0 & 1 \end{array} \right] \left[
\begin{array}{ccc}
1 & \frac{m_s}{m_d} |\overline{V}_{us}| &
\frac{m_b}{m_d} |V_{ub}| e^{i \delta} \\
0 & 1 & \frac{m_b}{m_s} |V_{cb}| \\
0 & 0 & 1 \end{array} \right] \mu_{\nu},
\end{equation}

\noindent We have scaled out an overall mass scale $\mu_{\nu}$ and parametrized the diagonal matrix $\overline{m}_{\nu}$ as ${\rm diag}(qe^{i \beta}, p e^{i \alpha}, 1)$.
There are nine neutrino observables: three masses, three MNS mixing angles, the Dirac CP-violating phase, and two Majorana CP-violating phases. These are determined by five model parameters, $p$, $q$, $\alpha$, $\beta$, and $\mu_{\nu}$. Therefore there are four predictions, which we may take to be $(\overline{M}_{\nu})_{ee}$ (which comes into neutrino-less double beta decay), and the three CP-violating phases. In \cite{BarrChen2012}, it is found that the model's best-fit values are $\delta_{Dirac} = 1.15 \pi$ radians, and 
$(\overline{M}_{\nu})_{ee} =$ 0.002 eV. 

But the model actually is considerably more predictive than counting parameters suggests, due to the fact that the expressions for observables in terms of model parameters are very non-linear. It is found that for good fits, certain quantities that have already been measured (such as, $\theta_{atm}$, $\theta_{sol}$, and $m_s/m_d$) must be in a restricted part of their present experimental range. For example, a value of the atmospheric angle smaller than $\pi/4$ is preferred by the fits, and a value of $m_s/m_d$ less than the median value of 20 is somewhat preferred. (See \cite{BarrChen2012} for details.)  As shown in \cite{BarrChen2013} the model also makes non-trivial predictions for branching ratios in proton decay, which we will not review here.

\section{Flavor Changing from Singlet Scalar Exchange} 

In this section we consider the effects of the scalar field ${\bf 1}_H$ that couples to the vector-like fermions to produces the $N \times N$ mass matrix $M_{AB} = 
Y'_{AB} \langle {\bf 1}_H \rangle$. We will henceforth call this singlet Higgs field 
$\Omega = \langle \Omega \rangle + \tilde{\Omega}$. The exchange of the $\tilde{\Omega}$ will mediate flavor-changing processes. For these effects to be observable in practice, we must assume that the scale $M_*$, which characterizes the mass and vacuum expectation value of $\Omega$, is not too much larger than the weak scale. We will assume that it is of order 1 to several TeV. 

Let us look first at the Yukawa couplings of $\tilde{\Omega}$ to the down-type quarks. In the same notation of Eq. (3), the Yukawa couplings of $\tilde{\Omega}$ to the down-type quarks is given by

\begin{equation}
\left( d_{({\bf 10})}, D_{({\bf 5}')} \right)  
\left( \begin{array}{cc} 0 & 0 \\ & \\
0 & M/\langle \Omega \rangle \end{array} \right) \left( \begin{array}{c} d^c_{(\overline{{\bf 5}})} \\ \\
D^c_{(\overline{{\bf 5}}')} \end{array} \right) \tilde{\Omega},
\end{equation}

\noindent When one block-diagonalizes to separate the light and heavy fermion stats, this Yukawa matrix is transformed by the unitary matrices $U_L$ and $U_R$ as in Eq. (4):

\begin{equation}
\left( \begin{array}{cc} 0 & 0 \\ & \\
0 & M/\langle \Omega \rangle \end{array} \right) \longrightarrow \underbrace{\left( \begin{array}{cc}
I & G^{\dag} \\ & \\ -G & I \end{array} \right)}_{\cong U_L^{\dag}} 
\left( \begin{array}{cc} 0 & 0 \\ & \\
0 & M/\langle \Omega \rangle \end{array} \right) \underbrace{\left( \begin{array}{cc} A & B \\ & \\
C & D \end{array} \right)}_{\cong U_R} = \frac{1}{\langle \Omega \rangle} \left( \begin{array}{cc} G^{\dag}MC & G^{\dag}MD  \\ & \\
MC & MD \end{array} \right). 
\end{equation}

\noindent So the effective Yukawa coupling of $\tilde{\Omega}$ to the three light 
down-type quarks $d$, $s$, and $b$, is given by $d_i (G^{\dag}MC)_{ij} d^c_j \; (\tilde{\Omega}/\langle \Omega \rangle)$. Remarkably, this Yukawa matrix, which we will call $Y_d$ can be written simply in terms of the master matrix $A$. Using Eq. (5), one gets

\begin{equation}
\begin{array}{ccl}
Y_d \langle \Omega \rangle \cong G^{\dag} M C & \cong & 
(- m_d T^{\dag} D^2 M^{-1}) M (- T A) \\
& = & m_d T^{\dag} D^2 T A \\
& = & m_d T^{\dag} (I + TT^{\dag})^{-1} T A \\
& = & m_d T^{\dag} T (I + T^{\dag} T)^{-1} A \\
& = & m_d (A^{-2} - I) A^3 = m_d(A - A A^{\dag} A).
\end{array}
\end{equation}

\noindent In going from line 3 to line 4, we have used the fact that $(I + T T^{\dag}T)^{-1} T = T (I + T^{\dag}T)^{-1}$, as can easily be seen by expanding out the 
expressions is parentheses as power series. In the last line, we have used the fact that $A$ is hermitian. Let us rewrite this expression in terms of the triangular matrix $A_{\Delta}$, since that is the matrix whose elements are known. Using Eq, (8) we have

\begin{equation}
\begin{array}{ccl}
Y_d \langle \Omega \rangle & = & m_d (A - A A^{\dag} A) \\
& = & m_d [{\cal D} A_{\Delta} {\cal U} -({\cal D} A_{\Delta} {\cal U})({\cal U}^{\dag} A_{\Delta}^{\dag} {\cal D}^*)({\cal D} A_{\Delta} {\cal U})] \\
& = & m_d {\cal D} A_{\Delta} \; [I - A_{\Delta}^{\dag} | {\cal D} |^2 A_{\Delta}] \; {\cal U}.
\end{array}
\end{equation}

\noindent The factor ${\cal U}$ on the right will be absorbed by the re-definition of the right-handed down-quark fields that was discussed after Eq. (9). Doing this re-definition, and using the fact that
$m_d {\cal D} A_{\Delta} = \overline{m}_d A_{\Delta} \equiv \overline{M}_d$, the Yukawa coupling matrix takes the form 

\begin{equation}
Y_d \langle \Omega \rangle = \overline{M}_d [I - A_{\Delta}^{\dag} | {\cal D} |^2 A_{\Delta}].
\end{equation}

\noindent The mass matrix $\overline{M}_d$ is diagonalized by a bi-unitary transformation to give 
$V_L^{\dag} \overline{M}_d V_R = M_d^{phys} = {\rm diag} (m_d, m_s, m_b)$. From Eq. (11). One sees that the matrix $V_L$ is the CKM matrix, while the matrix $V_R$ differs from the identity matrix by terms of order $O(m_s^2/m_b^2)$, $O(m_d m_s/m_b^2)$, and $O(m_d^2/m_b^2)$, which can be neglected. It is clear then that in the physical basis of the down quarks

\begin{equation}
Y_d^{phys} \cong \frac{1}{\langle \Omega \rangle} M_d^{phys} [I - A_{\Delta}^{\dag} | {\cal D} |^2 A_{\Delta}].
\end{equation}

\noindent Obviously, only the second term in the brackets leads to flavor changing. Let us parametrize the unknown matrix ${\cal D}$ as diag$(\delta, \epsilon, \zeta)$.
The flavor-changing Yukawa coupling matrix of the $\tilde{\Omega}$ to the physical 
down-type quarks is of the form
$d_i (Y_d^{FC})_{ij} d^c_j \tilde{\Omega}$, where

\begin{equation} 
\begin{array}{ccl} 
Y_d^{FC} & = & \frac{-1}{\langle \Omega \rangle}
\left( \begin{array}{ccc} m_d & 0 & 0 \\
0 & m_s & 0 \\ 0 & 0 & m_b \end{array} \right)    
 \left( \begin{array}{ccc} 1 & 0 & 0 \\
b & 1 & 0 \\ ce^{-i \theta} & a & 1 \end{array} \right)
\left( \begin{array}{ccc} |\delta|^2 & 0 & 0 \\
0 & |\epsilon|^2 & 0 \\ 0 & 0 & |\zeta|^2 \end{array} \right) 
\left( \begin{array}{ccc} 1 & b & c e^{i \theta} \\ 0 & 1 & a \\ 0 & 0 & 1 \end{array}
\right) \\ & & \\
& = & \frac{-1}{\langle \Omega \rangle} \left( \begin{array}{ccc} m_d & 0 & 0 \\
0 & m_s & 0 \\ 0 & 0 & m_b \end{array} \right) \left(
\begin{array}{ccc} \Delta_{dd} & \Delta_{ds} & \Delta_{db} \\
\Delta_{sd} & \Delta_{ss} & \Delta_{sb} \\
\Delta_{bd} & \Delta_{bs} & \Delta_{bb} \end{array}
\right), 
\end{array}
\end{equation}

\noindent where

\begin{equation}
\begin{array}{ccl}
\Delta_{d s} & = & \Delta_{sd} = |\delta|^2 b, \\ 
\Delta_{d b} & = & \Delta_{bd}^* = |\delta|^2 c e^{i \theta}, \\
\Delta_{sb} & = & \Delta_{bs}^* = |\epsilon|^2 a + |\delta|^2 bc e^{i \theta}.
\end{array}
\end{equation}

\noindent Note that the flavor-changing (i.e. off-diagonal) elements of
$Y_d^{FC}$ depend only on two unknown combinations of parameters:
$|\delta|^2/\langle \Omega \rangle$ and $|\epsilon|^2/\langle \Omega \rangle$.
Note also that $\Delta_{d s}$ and $\Delta_{sd}$ are real in the physical basis of the 
quarks, so that the $\epsilon$ parameter of the $K^0 - \overline{K^0}$ system does not put constraints on flavor changing coming from the singlet scalar exchange.

The charged-lepton sector is identical except for a left-right transposition.  So writing the flavor-changing Yukawa coupling matrix of the $\tilde{\Omega}$ to the physical 
charged leptons as
$\ell^+_i (Y_{\ell}^{FC})_{ij} \ell^-_j \tilde{\Omega}$, one finds

\begin{equation} 
\begin{array}{ccl} 
Y_{\ell}^{FC} & = & \frac{1}{\langle \Omega \rangle}
\left( \begin{array}{ccc} m_e & 0 & 0 \\
0 & m_{\mu} & 0 \\ 0 & 0 & m_{\tau} \end{array} \right)    
 \left( \begin{array}{ccc} 1 & 0 & 0 \\
b & 1 & 0 \\ ce^{-i \theta} & a & 1 \end{array} \right)
\left( \begin{array}{ccc} |\delta|^2 & 0 & 0 \\
0 & |\epsilon|^2 & 0 \\ 0 & 0 & |\zeta|^2 \end{array} \right) 
\left( \begin{array}{ccc} 1 & b & c e^{i \theta} \\ 0 & 1 & a \\ 0 & 0 & 1 \end{array}
\right) \\ & & \\
& = & \frac{1}{\langle \Omega \rangle} \left( \begin{array}{ccc} m_e & 0 & 0 \\
0 & m_{\mu} & 0 \\ 0 & 0 & m_{\tau} \end{array} \right) \left(
\begin{array}{ccc} \Delta_{ee} & \Delta_{e \mu} & \Delta_{e \tau} \\
\Delta_{\mu e} & \Delta_{\mu \mu} & \Delta_{\mu \tau} \\
\Delta_{\tau e} & \Delta_{\tau \mu} & \Delta_{\tau \tau} \end{array}
\right), 
\end{array}
\end{equation}

\noindent where

\begin{equation}
\begin{array}{ccl}
\Delta_{e \mu} & = & \Delta_{\mu e} = |\delta|^2 b, \\ 
\Delta_{e \tau} & = & \Delta_{\tau e}^* = |\delta|^2 c e^{i \theta}, \\
\Delta_{\mu \tau} & = & \Delta_{\tau \mu}^* = |\epsilon|^2 a + |\delta|^2 bc e^{i \theta}.
\end{array}
\end{equation}

The flavor-changing Yukawa couplings come into the processes 
$\ell_1 \rightarrow \ell_2 \gamma$ through-two loop diagrams, as shown in \cite{legamma}. The specific diagrams that dominate in this model have the vector-like fermions running around the loop that gives an effective $\tilde{\Omega}$-photon-photon coupling. The resulting branching ratios for the flavor-changing lepton decays can be expressed in terms of the quantities given in Eq. (24) as follows 
\cite{dchang}:

\begin{equation}
BR(\ell_1 \rightarrow \ell_2 \gamma)  \cong  24 \left( \frac{\alpha}{\pi} \right)^3
\left( \frac{v}{\langle \Omega \rangle} \right)^4 |\Delta_{\ell_1 \ell_2} |^2.
\end{equation}

\noindent One prediction is that

\begin{equation}
\frac{BR(\tau \rightarrow e \gamma)}{BR(\mu \rightarrow e \gamma)} 
\cong \left| \frac{\Delta_{e \tau}}{\Delta_{e \mu}} \right|^2 = \left| \frac{c}{b} \right|^2 = \left( \frac{m_b}{m_s} \right)^2  \left| \frac{V_{ub}}{V_{us}} \right|^2
\approx 1.
\end{equation}

\noindent If one assumes that the expression for $\Delta_{\mu \tau}$ in Eq. (24) is dominated by the $|\delta|^2$ term, then one would also have the prediction

\begin{equation} 
BR(\tau \rightarrow \mu \gamma) \cong |c|^2 BR(\mu \rightarrow e \gamma)
\cong 16 \cdot BR(\mu \rightarrow e \gamma).
\end{equation} 

\noindent Given the present limit \cite{meg} that $BR(\mu \rightarrow e \gamma) <
5.7 \times 10^{-13}$, this would gives a prediction that 
$BR(\tau \rightarrow \mu \gamma) < 10^{-11}$. This is well below even what is expected to be observable at a super-$c$-$\tau$ factory \cite{ctaufactory}. On the other hand, the branching ratio for this decay can be much larger if $\Delta_{\mu \tau}$ in Eq. (23) is dominated by the $|\epsilon|^2$ term. As we will show below, there is an approximate theoretical bound that $|\epsilon|^2 < 1/2$. This would give

\begin{equation}
BR(\tau \rightarrow \mu \gamma) \leq 1.5 \times 10^{-9} \left( 
\frac{1 \; {\rm TeV}}{\langle \Omega \rangle} \right)^4.
\end{equation}

The flavor-changing processes involving quarks do not get large enough contributions from
the exchange of the singlet scalar $\tilde{\Omega}$ to stand out from Standard Model contributions. For instance, the coefficient of $(\overline{s} d)(\overline{s} d)$ operators is found from from Eqs. (23-24) to be of order $  \frac{m_s^2}{\langle \Omega \rangle^2 M_{\tilde{\Omega}}^2} b^2 |\delta|^4 < 10^{-15} ({\rm GeV})^{-2} \left( \frac{1\; {\rm TeV}}{M_*} \right)^4$, where we have used an upper bound on $|\delta|^2$ that is derived below. (From the first line of Eq. (32) one finds that $|\delta|^2$ must be less than $(1 + b^2)^{-1} \sim 1/17$.) 

Let us now consider the parameters $\delta, \epsilon, \zeta$. While the matrix ${\cal D} = {\rm diag} (\delta, \epsilon, \zeta)$
is not known {\it a priori}, it is nevertheless possible to
derive strict upper bounds on the parameters $|\delta|$, $|\epsilon|$,
and $|\zeta|$ from the properties of the master matrix $A$.
From the fact that $A \equiv
(I + T^{\dag} T)^{-1/2}$ and that $A ={\cal D} A_{\Delta} {\cal U}$, one has that

\begin{equation}
\begin{array}{l}
A \; A^{\dag}  =  {\cal D} A_{\Delta} A_{\Delta}^{\dag} {\cal D}^*
= (I + T^{\dag} T)^{-1} \\  \\
({\cal D} A_{\Delta} A_{\Delta}^{\dag} {\cal D}^*)^{-1} - I =
T^{\dag} T.
\end{array}
\end{equation}

\noindent Computing the matrix on the left side of the above equation, one obtains

\begin{equation}
\left[ \begin{array}{ccc}
1/|\delta|^2 & -b/(\delta^* \epsilon) &
(ab- c e^{i \theta})/(\delta^* \zeta) \\
-b/(\delta \epsilon^*) & (1 + b^2)/|\epsilon|^2
 & -(a + ab^2 -bce^{i \theta})/(\epsilon^* \zeta) \\
(ab - c e^{-i \theta})/(\delta \zeta^*) &
-(a + ab^2 -bce^{-i \theta})/(\epsilon \zeta^*) &
(1 + a^2 + |ab - c e^{i \theta}|^2)/|\zeta|^2 \end{array}
\right] - I = T^{\dag} T.
\end{equation}

\noindent For any matrix $T$, there is an inequality that must
be satisfied by the elements of $T^{\dag} T$. namely

\begin{equation}
|(T^{\dag}T)_{ij}|^2 \leq (T^{\dag}T)_{ii} (T^{\dag}T)_{jj},
\;\;\; \forall \;\; i,j.
\end{equation}

\noindent This is obvious if we write $T_{ij} = (\vec{t}_{(j)})_i$, where
$\vec{t}_{(i)}, i = 1,2,3$, are three complex vectors.
Then the inequality is just seen to be the statement
that $|\vec{t}_{(i)}^* \cdot \vec{t}_{(j)}| \leq
|\vec{t}_{(i)}|\; |\vec{t}_{(j)}|$. From this inequality
with $(i,j) = (1,2)$, $(1,3)$, and $(2,3)$, respectively,
one obtains after a little algebra

\begin{equation}
\begin{array}{l}
(1 + b^2) |\delta|^2 + |\epsilon|^2 \;\; \leq  \;\; 1, \\ \\
(1 + a^2 +|ab - c e^{i \theta}|^2) |\delta|^2 + |\zeta|^2
\;\; \leq  \;\; 1 + a^2, \\ \\
(1 + a^2 +|ab - c e^{i \theta}|^2) |\epsilon|^2 +
(1 + b^2) |\zeta|^2
\;\; \leq  \;\; 1 + b^2 + c^2.
\end{array}
\end{equation}

\noindent using the values of $a$, $b$, $c$ and $\theta$ given in Eq. (12), the third equation of Eq. (32) gives an upper bound on $|\epsilon|^2$ of approximately $1/2$, as used in Eq. (28).

In this paper, we have assumed that the scale $M_*$ of $\langle \Omega \rangle$ is in the low TeV range, as otherwise the flavor-changing effects from exchanges of $\tilde{\Omega}$ would be hopelessly small. But then one must run the Yukawa couplings $Y'_{AB}$ and $y'_{Aj}$ shown in the last line of Eq. (1) from the GUT scale down to the scale $M_*$. If these ran differently for the leptons and quarks, it would make the matrices $\Delta$ and $M$ in Eq. (3) different for 
quarks and leptons, and thus also make the master matrix $A$ different for quarks and leptons. That could destroy the predictivity of the model. If one considered only gluon loops in the running there is no problem, as the effect would be to increase $\Delta$ and $M$ by the same factor for quarks relative to leptons. This factor would cancel in the ratio $T = M^{-1} \Delta$, and therefore also in $A = [I + T^{\dag} T]^{-1/2}$.
However, the gluon loops could do the following: they could increase the Yukawa couplings $Y'_{AB}$ and $y'_{Aj}$ for quarks to such an extent that the effect of these Yukawas on their own running could be much more significant for quarks than for leptons. That would make the forms of the matrices $\Delta$ and $M$ --- and therefore the form of $A$ --- different for quarks and leptons. 

There are two ways to avoid this problem. One is that all the Yukawas $Y'_{AB}$ and $y'_{Ai}$ remained small for the whole range from $M_{GUT}$ to $M_*$. This has a drawback, however. If these Yukawa couplings $Y'_{AB}$ are small compared to 1, then the VEV
$\langle \Omega \rangle$ would have to be large compared to a TeV to make the vector-like
fermions in ${\bf 5}' + \overline{{\bf 5}}'$ heavy enough not to be seen. That would suppress flavor-changing effects from $\tilde{\Omega}$ exchange.

A cleaner way to avoid the problem is to assume the following two conditions: (a) The Yukawa couplings $y'_{Ai}$ that generate the mass matrix $\Delta$ are small compared to 1, and the VEV of the Higgs fields ${\bf 1}'_{Ai}$ correspondingly large compared to a TeV.
(That would have the additional advantage of making flavor changing from the exchange of these scalars negligible.) (b) The Yukawa coupling matrix
$Y'_{AB}$ is proportional to the identity matrix, which could be the result of a flavor symmetry that acted on the vector-like families. Then even if gluon loops drove $Y'_{AB}$ to be of order 1 at low scales, that would not affect the form of $Y'_{AB}$.  
 
Another theoretical issue raised by $M_*$ being near the weak scale is that 
the spontaneous breaking of the family symmetry group $G_F$ would cause cosmological domain walls. This breaking is done by the VEVs $\langle {\bf 1}_{Hi} \rangle$. To avoid overclosing the universe, these domain walls would have to be ``inflated away". One simple possibility is that $G_F$ is actually broken at a scale much higher than $M_*$ but
only induces a VEV for ${\bf 1}_{Hi}$ that is of order $M_*$. For example, consider the terms ${\cal L}(\sigma) 
= - \frac{1}{2} M^2 \sigma^2 + \overline{\psi} \psi \sigma$, where the scalar field 
$\sigma$ and fermion bilinear $\overline{\psi} \psi$ are odd under a $Z_2$ and $M$ is of order the GUT scale. Let the fermion bilinear get a condensate $\langle \overline{\psi} \psi \rangle = \Lambda^3$, where $\Lambda \sim (M^2 M_*)^{1/3}$, which is many orders of magnitude bigger than $M_*$. The $Z_2$ will be broken at the scale $\Lambda$, whereas  $\langle \sigma \rangle = \Lambda^3/M^2 \sim M_*$.

\section{Conclusions}

The model of flavor symmetry and flavor violation proposed in \cite{BarrChen2012} has the virtue that it is (a) conceptually simple, (b) explains some of the qualitative features 
of the quark and lepton spectrum ({\it e.g.} the MNS angles being much larger than the CKM angles), and (c) is highly predictive. As such, it can provide a kind of ``benchmark" for seeing how large various kinds of flavor-changing processes might be expected to be. 

The model is of the ``lopsided" type \cite{lopsided}, which tends to give relatively large flavor-changing effects.
In models with symmetric mass matrices, which are very common in the literature, off-diagonal Yukawa couplings $Y_{ij}$ are typically proportional to
$\sqrt{m_i m_j}/v$. In lopsided models, however, $Y_{ij}$ and $Y_{ji}$ are very different in magnitude from each other, being proportional to $m_i/v$ and $m_j/v$. This is the reason for the name ``lopsided", and why the flavor-changing effects tend to be relatively large. 

It is likely, then, that the flavor-changing Yukawa couplings given in Eqs. (21-24) 
(with the bounds in Eq. (32)) are typical of what would expect for a new scalar field. We see that if the scale of new physics $M_*$ is of order 1 TeV, there is good hope of eventually seeing the processes $\tau \rightarrow \mu \gamma$, 
$\tau \rightarrow e \gamma$, and 
$\mu \rightarrow e \gamma$. One also sees from this model, that observing such processes can confirm or rule out specific models of the origin of flavor and flavor violation.


\begin{thebibliography}{999}
\bibitem{BarrChen2012} S.M. Barr and H.Y. Chen, {\it JHEP} {\bf 1211}, 92 (2012). 
\bibitem{BarrChen2013} S.M. Barr and H.Y. Chen, {\it JHEP} {\bf 1310}, 49 (2012). 
\bibitem{MNS} Z. Maki, M. Nakagawa and S. Sakata, {\it Prog. Theor.
Phys.} {\bf 28}, 870 (1962).
\bibitem{CKM} N. Cabibbo, {\it Phys. Rev. Lett.} {\bf 10}, 531
(1963); M. Kobayashi and T. Maskawa,{\it Prog. Theor. Phys.} {\bf
49}, 652 (1973).
\bibitem{lopsided} K.S. Babu and S.M. Barr, {\it Phys. Lett.} {\bf B381}, 202 (1996);
C.H. Albright and S.M. Barr, {\it Phys. Rev.} {\bf D58}, 013002 (1998);
J. Sato and T. Yanagida, {\it Phys. Lett.} {\bf B430}, 127 (1998);
C.H. Albright, K.S. Babu and S.M. Barr, {\it Phys. Rev. Lett.} {\bf
81} 1167 (1998); N. Irges, S. Lavignac, and P. Ramond, {\it Phys.
Rev.} {\bf D58}, 035003 (1998); K.S. Babu, J.C. Pati and F. Wilczek,
{\it Nucl. Phys.} {\bf B566}, 33 92000); J. Sato and T. Yanagida,
{\it Phys. Lett.} {\bf B493}, 356 (2000); T. Asaka, {\it Phys.
Lett.} {\bf B562}, 791 (2003); X.D. Ji, Y.C. Li, and R.N. Mohapatra,
{\it Phys. Lett.} {\bf B633}, 755 (2006).
\bibitem{legamma} J.D. Bjorken and S. Weinberg, {\it Phys. Rev. Lett.} {\bf 38},
622 (1977); S.M. Barr and A. Zee, {\it Phys. Rev. Lett.} {\bf 65}, 21 (1990).
\bibitem{dchang} D. Chang, W.S. Hou, and W.-Y. Keung, {\it Phys. Rev.} {\bf D48}, 217 (1993).
\bibitem{meg} Gianluca Cavoto, in 12th conference on Flavor Physics and CP Violation (FPCP 2014), Marseille, France, arxiv:1407.8327[hep-ex]. 
\bibitem{ctaufactory} A.V. Bobrov and A.E Bondar, {\it Nucl. Phys. Proc. Suppl.} {\bf 225-227} 195-197 (2012). 
\end{thebibliography}
\end{document}